\documentclass{article}
\usepackage{icrctc07}
\pdfoutput=1

\title{Cherenkov Telescope Array: The next-generation ground-based gamma-ray observatory}
\shorttitle{Cherenkov Telescope Array}
\authors{G. Hermann$^{1}$,W. Hofmann$^{1}$, T. Schweizer$^{2}$, M. Teshima$^{2}$ for the CTA consortium}
\shortauthors{G. Hermann and et al}
\afiliations{$^1$Max Planck Institute for Nuclear Physics, Saupfercheckweg 1, 69117 Heidelberg,  
Germany\\$^2$Max Planck Institute for Physics (Werner-Heisenberg-Institut), F\"ohringer Ring 6, 80805 Munich, Germany}
\email{tschweiz@mppmu.mpg.de}

\abstract{High energy gamma-ray astronomy is a newly emerging and very successful branch of astronomy and astrophysics. 
Exciting results have been obtained by the current generation Cherenkov telescope systems such as H.E.S.S., 
MAGIC, VERITAS and CANGAROO. The H.E.S.S. survey of the galactic plane has revealed a large number of sources 
and addresses issues such as the question about the origin of cosmic rays. 
The detection of very high 
energy emission from extragalactic sources at large distances has provided insights in the star formation during 
the history of the universe and in the understanding of active galactic nuclei. The development of the very large 
Cherenkov telescope array system (CTA) with a sensitivity about an order of magnitude better than current 
instruments and significantly improved sensitivity is under intense discussion. This observatory will reveal an order of magnitude 
more sources and due to its higher sensitivity and angular resolution it will be able to detect new classes of 
objects and phenomena that have not been visible until now. A combination of different telescope types will 
provide the sensitivity needed in different energy ranges. }

\begin{document}
\maketitle

\section{Introduction}
The Cherenkov Telescope Array (CTA) \cite{ref_CTA} is a proposed advanced facility for ground 
based high-energy gamma ray astronomy, 
based on the observation of Cherenkov radiation. This approach has proven to be 
extremely successful for gamma rays of energies above a 100~GeV. The facility will 
consist of an array of telescopes enhancing the all sky monitoring capability.
CTA is designed to be the next generation observatory after currently running Cherenkov telescope projects
such as the H.E.S.S. experiment \cite{ref_HESS}, MAGIC \cite{ref_MAGIC}, VERITAS \cite{ref_VERITAS}
and CANGAROO \cite{ref_CANGAROO}. In order to cover the full sky, it is 
planned to build two stations, one in the northern hemisphere and the other one in the southern hemisphere.

\begin{figure}
\begin{center}
\includegraphics [width=0.55\textwidth]{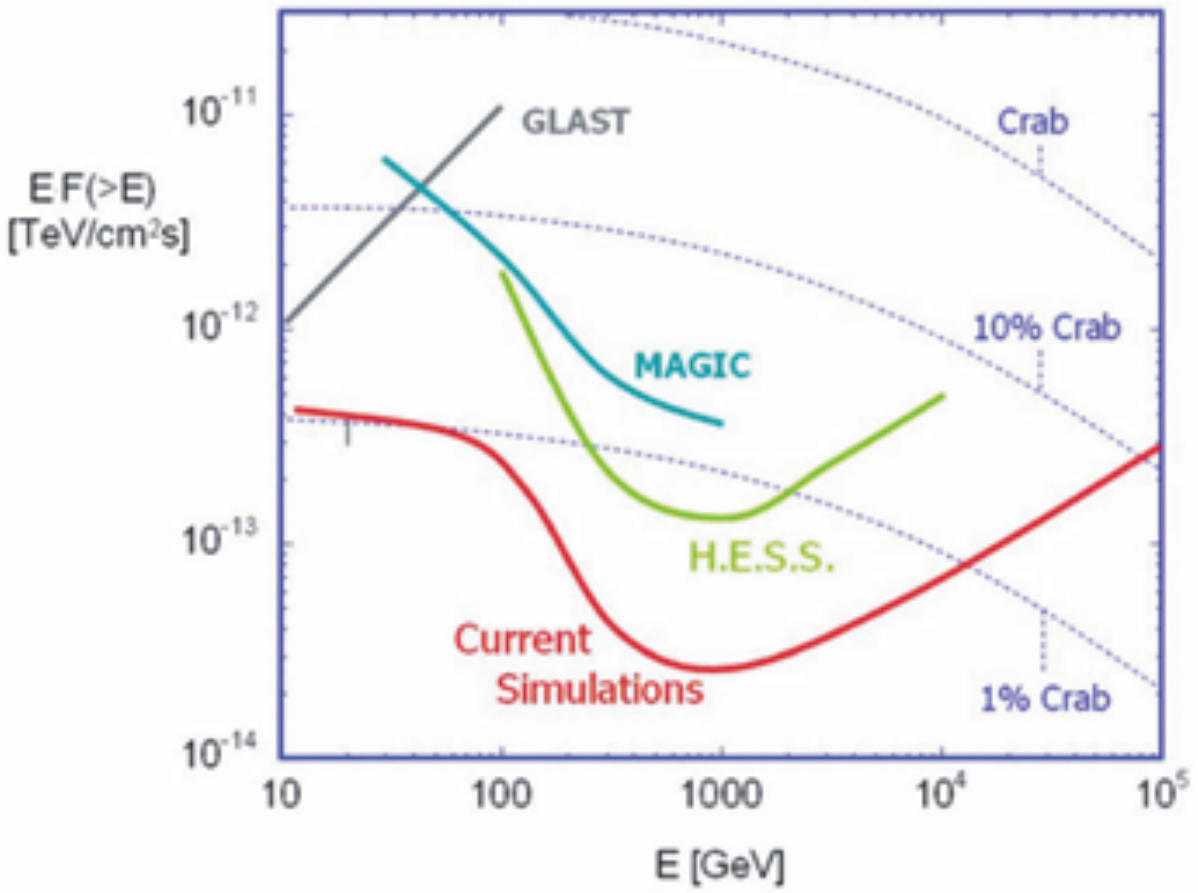}
\end{center}
\caption{The figure illustrates the sensitivity that is aimed at with CTA. The exact values will 
finally depend on the  real layout of the system.}\label{fig1}
\end{figure}

CTA will be about a factor of ten more sensitive than
current experiments (see Fig.\ref{fig1}). It will allow in-depth studies of known classes of gamma emitters and 
also detect new source classes that are below the sensitivity of current instruments. In its core energy range
from about 100~GeV up to 10~TeV, CTA will reach millicrab sensitivity. The full energy range 
will be three to four orders of magnitude from some 10~GeV up to 100~TeV, which is crucial
for the understanding of the physical processes. Together with multiwavelength observations from radio wavelengths to
optical, X-ray and MeV energy range, CTA will reveal deep insight in the most violent and energetic physical processes
in the universe. 
For selected sets of gamma-ray events, CTA will have an angular resolution by a factor of
5 better than current instruments because of the large amount of telescopes within the array. 
This angular resolution will be necessary for resolving the high density 
of galactic objects within the milky way. It allows also detailed morphological studies of extended objects.
Due to its large detection area and lower threshold it will collect a much larger event statistics and
improve the time resolution of transient objects down to the sub-minute scale.
Some few telescopes can be used for monitoring transient objects. This will deliver interesting 
details about the acceleration process of particles  in cosmic accelerators. All in all, 
CTA is expected to detect and observe about 1000 sources.

In March 2007 a European FP7 design study was applied for to examine the exact system layout and
technical details of the CTA. 34 institutes in 15 countries (France, Germany, Italy, Spain, United Kingdom, Poland, 
Finnland, Switzerland, Netherlands, Czech Republic, Armenia, Ireland, United States, Republic of South Africa)
are participating in the CTA design study. 

\section{Physics prospects of CTA}
The large list of the astrophysics cases for CTA can be roughly split into two parts, the galactic sources and 
the extragalactic sources. In this short overview we will only mention the most promising science cases.

\begin{figure}
\begin{center}
\includegraphics [width=0.48\textwidth]{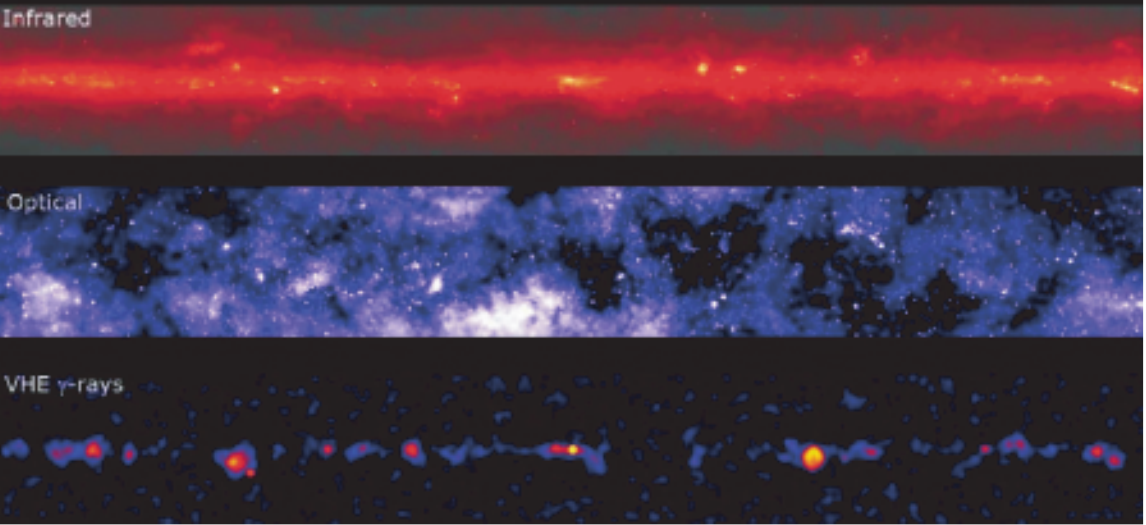}
\end{center}
\caption{The milky way seen in multiwavelength observations.The H.E.S.S. observations at TeV energies reveal a 
plethora of  sources. Most of them are supernova remnants and pulsar wind nebulae.}\label{fig2}
\end{figure}

In the galaxy many TeV gamma sources have already been discovered, mainly in the galactic scan by 
H.E.S.S. (see Fig.\ref{fig2}). The sources are supernova remnants (SNRs), pulsar wind nebulae (PWNs), binary systems,
the galactic center itself and diffuse gamma 
radiation from our galaxy. The galactic astrophysical
subjects are manyfold. One main subject is the question about the origin of cosmic rays (CR).

The origin of the bulk of CR observed at earth remains unknown even almost one century after
their discovery by Victor Hess in 1912. Several arguments hint to supernova remnants (SNRs)  
being the main candidate sources, but a solid evidence is still 
missing. The detection of TeV radiation from 
30-100 SNRs that will be detected by an instrument with a higher sensitivity 
would be the ultimate proof of these sources to be the bulk producers of CR arriving
at earth. The measurement of their spectra at 100~GeV up to 100~TeV region will reveal the
physics of the particle acceleration mechanism and probe the 'knee' in the hadronic spectrum of CR
at $3\cdot10^{15}eV$ using the gamma emission radiated during the acceleration process of particles at the 
shock front of the SNR with the interstellar medium. The diffuse gamma radiation from the region around the
galactic center  seems to originate from the interaction of cosmic rays with molecular clouds.
Molecular clouds are a tool that allow to probe the cosmic ray spectrum in different locations in the galaxy.
A new instrument like CTA has the necessary sensitivity to improve these measurements and even
separate additional contributions to diffuse radiation due to more exotic physics.

In addition, also pulsar wind nebulae  as well as binary systems might accelerate not only leptons
but significantly also hadrons to multihundred TeV radiation and therefore 
contribute to the CRs. Purely leptonic acceleration models as well as hadronic acceleration models have been 
suggested for both cases. Precise multiwavelength
measurements from radio, optical, X-ray, MeV up to 100 TeV gamma radiation enable phycisists
to fit their models to the full wavelength range and verify or falsify one or another of these models.
It is therefore crucial to measure an  energy range as wide as possible from several tens GeV up to
100 TeV. CTA will provide this energy range.

Very interesting source classes are the various types of binary systems, Wolf-Rayet-Star binaries, Be-Star-Pulsar
binary, Star-black hole binary, micro quasars etc. which are transient TeV gamma emitters. They are
ideal objects to study various types of particle acceleration and Gamma ray absorption.

The low energy threshold together with a good sensitivity will allow to detect many extragalactic
objects at high redshifts that are not visible at higher energies due to the so-called gamma ray horizon. 
High energy gamma rays are absorbed due to electron-positron pair creation with extragalactic background 
light in the universe. 
The study of the spectra and the  energy cut-offs of active galactic nuclei (AGN) enables us to calculate 
and measure the spectrum of the so-called extragalactic background light (EBL). 
The main contributions to the EBL is redshifted starlight and its re-emission from dust 
in the universe. This measurement not only reflects 
the history of star formation in the universe but also contains information about the evolution 
of the universe expressed in the Hubble constant and cosmic constant.

Simultaneous multi wavelength campains of the transient AGN with instruments at other wavelengths will 
supply more information about leptonic and hadronic acceleration models in AGN jets by observing these 
objects in the multiwavelength band, same as for the galactic sources. It is not understood if
the highest energy CR have been accelerated by these objects.

The final proof of hadronic acceleration can be only given by the detection of
a neutrino signal from these sources. Therefore, multimessenger observations 
are important in gamma astronomy. The northern CTA observatory will observe the same hemisphere 
of the sky as the ICECUBE neutrino telescope. The future KM3NET neutrino telescope will 
observe the southern sky and will probably start operation around 2015. 

AGN exhibit very fast flux changes at all wavelengths that are especially difficult to explain for
the TeV radiation. The large scales of the emission regions in the jet imply large gamma factors for
the bulk motion. 

Another type of very interesting transient objects are the so-called gamma ray bursts. These objects 
exhibit ultrafast X-ray and gamma-ray flux variability on second and subsecond time scale. As of today, 
no radiation at hundreds of GeV energies could be detected, it is believed mainly due to the gamma ray horizon. 

Very distant transient objects with ultrafast flux variability can be used to test Lorentz invariance
violation.

Furthermore, CTA has considerable discovery potential e.g. for the detection 
additional non-blazar AGN  of pulsed gamma emission from pulsars, 
clusters of galaxies and the detection of a dark matter candidate, the neutralino.

\section{Design of the CTA system}
\begin{figure}
\begin{center}
\includegraphics [width=0.48\textwidth]{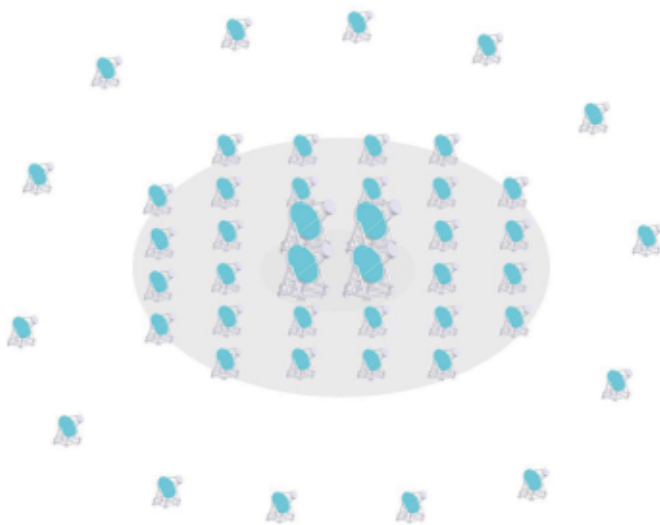}
\end{center}
\caption{This figure illustrates possible telescope configurations which achieve the aimed sensitivity.}\label{fig3}
\end{figure}

The exact design of CTA is being studied and not precisely defined as of now. First MC studies indicate some possible
scenarios in which the sensitivity and angular resolution aimed for can be achieved. Fig.~\ref{fig3}
illustrates the possible design, a combination of arrays of 2-3 different telescope sizes. A large
number (several tens) of mid-size telscopes will provide for the millicrab sensitivity and high angular 
resolution in the core energy range from about 100 GeV up to 10 TeV. An extension of a few (4-9) very large 
diameter telescopes and many (up to 100) very small telescopes distributed over a large area enlarge the energy 
range from several 10 GeV up to 100 TeV.

In this way one achieves an energy coverage of 3-4 orders of magnitude while keeping the high sensitivity.
This is crucial for detailed studies of the shape of the spectrum which contains a lot of information about
the mechanism of particle acceleration and physics in the object. 

In order to achieve full sky coverage it is planned to install two stations, one in the southern hemisphere
and one in the northern hemisphere.

The design of the telescope camera and the performance predictions of CTA will be based on 
classical photomultipliers but an eye is kept also on the technical development of future photosensors 
(such as SiPM) with higher photon detection efficiency to lower the energy threshold of the system.

Detailed MC simulations will define the details such as FOV, optimal telescope sizes, 
camera pixel sizes and telescope spacing in the array.

The real challenge will be to reach the perfection and reliability of the hardware that is needed to run
a large number of telescopes without major interruptions or failure in a quasi robotic modus since it
is simply impossible to do frequent repairs on every telescope.

The design should start in 2008 and continue until the end of 2011. We aim to begin the construction 
of the full array around 2012. It is intented to have an overlap with the GLAST satellite project.

\section{Conclusions}
We present the new Cherenkov Telescope Array (CTA) project that unifies for the first time the 
research groups working in gamma ray astronomy in a common strategy, resulting 
in a new facility that is well beyond possible upgrades of
existing instruments like H.E.S.S., MAGIC, VERITAS and CANGAROO. The new facility will be run as 
an observatory open to external astronomers. Its sensitivity will be better
by about one order of magnitude combined with a larger energy range coverage. It is expected that 
at the order of 1000 new sources and also new classes of sources will be discovered 
which have been not visible with today's instruments.

The physics of most objects can only be understood, if the whole multi-wavelength picture is 
generated. One of the main topics of CTA will be therefore the multi-wavelength and multi-messenger 
observation together with X-ray and gamma-ray satellites,  optical and radio telescopes as well 
as with the forthcoming neutrino experiments.

\section{Acknowledgements}
We wish to thank all the scientists in 34 institutes in 15 countries who are contributing and 
participating in the Cherenkov Telescope Array (CTA) project. (See http://www.mpi-hd.mpg.de/hfm/CTA/)


\begin{thebibliography}{1}

\bibitem{ref_CTA}
"{The Cherenkov Telescope Project}".
\newblock {\em \\http://www.mpi-hd.mpg.de/hfm/CTA/}.

\bibitem{ref_HESS}
"{The H.E.S.S. project}".
\newblock {\em \\http://www.mpi-hd.mpg.de/hfm/HESS/}.

\bibitem{ref_MAGIC}
"{The MAGIC Telescope}".
\newblock {\em \\http://wwwmagic.mppmu.mpg.de/}.

\bibitem{ref_VERITAS}
"{The VERITAS project}".
\newblock {\em \\http://veritas.sao.arizona.edu/}.

\bibitem{ref_CANGAROO}
"{The CANGAROO telescope system}".
\newblock {\em \\http://icrhp9.icrr.u-tokyo.ac.jp/}.


\end{thebibliography}

\end{document}